\documentclass[a4paper,12pt]{article}

\usepackage{amsmath,amssymb,amsfonts}
\usepackage{epsfig,cite, cancel, slashed}
\usepackage[footnotesize]{caption}

\pdfoutput=1

\newlength{\xtrawidth}
\newlength{\xtraheight}
\newcommand{\cref}[1]{Chapter~\ref{#1}}
\newcommand{\bcenter}{\begin{center}}
\newcommand{\ecenter}{\end{center}}
\newcommand{\beq}{\begin{equation}}
\newcommand{\eeq}{\end{equation}}
\newcommand{\bea}{\begin{eqnarray}}
\newcommand{\eea}{\end{eqnarray}}
\newcommand{\bean}{\begin{eqnarray*}}
\newcommand{\eean}{\end{eqnarray*}}
\newcommand{\ba}{\begin{array}}
\newcommand{\ea}{\end{array}}
\newcommand{\ben}{\begin{enumerate}}
\newcommand{\een}{\end{enumerate}}
\newcommand{\bi}{\begin{itemize}}
\newcommand{\ei}{\end{itemize}}
\newcommand{\bd}{\begin{description}}
\newcommand{\ed}{\end{description}}
\newcommand{\bdiag}{\begin{diagram}}
\newcommand{\ediag}{\end{diagram}}
\def\fnote#1#2{\begingroup\def\thefootnote{#1}\footnote{#2}
     \addtocounter{footnote}{-1}\endgroup}

\def\IC{\mathbb{C}}

\def\IR{\mathbb{R}}
\def\IP{\mathbb{P}}
\def\IZ{\mathbb{Z}}
\def\cO{{\mathcal O}}

\newcommand{\nn}{\nonumber}

\newcommand{\cA}{{\cal A}}

\newcommand\cN{{\cal N}}
\newcommand\cG{{\cal G}}
\newcommand\cH{{\cal H}}

\newcommand{\be}{\begin{equation}}
\newcommand{\ee}{\end{equation}}

\newcommand{\comment}[1]{}

\def\IC{\mathbb{C}}

\def\IR{\mathbb{R}}
\def\IP{\mathbb{P}}
\def\IZ{\mathbb{Z}}

\numberwithin{equation}{section}

\begin{document}


\title{\LARGE \bf{Combinatorics in $\cN=1$ Heterotic Vacua }}
\author{
Seung-Joo Lee
}
\date{}
\maketitle
\begin{center}
{\small
{\it School of Physics, Korea Institute for Advanced Study, Seoul 130-722, Korea}
\fnote{}{s.lee@kias.re.kr}
}
\end{center}

\abstract{We briefly review an algorithmic strategy to explore the landscape of heterotic $E_8 \times E_8$ vacua, in the context of compactifying smooth Calabi-Yau three-folds with vector bundles. The Calabi-Yau three-folds are algebraically realised as hypersurfaces in toric varieties and a large class of vector bundles are constructed thereon as monads. In the spirit of searching for Standard-like heterotic vacua, emphasis is placed on the integer combinatorics of the model--building programme. 
}


\section{Introduction}
Compactifications of $E_8 \times E_8$ heterotic theory \cite{Candelas:1985en, GSW} and heterotic M-theory \cite{Witten:1996mz, Lukas:1997fg, Lukas:1998yy, Lukas:1998hk, Lukas:1998tt} on smooth Calabi-Yau three-folds is a simple and compelling way to reach $\cN=1$ supersymmetry at four dimensions. 
A Calabi-Yau three-fold necessarily admits a Ricci--flat metric $g_{\alpha \bar \beta}$, where $\alpha, \bar \beta = 1,2,3$ are respectively, the holomorphic and anti-holomorphic indices. 
One also turns on an internal gauge field, in a sub-algebra $\cG$ of the full $E_8$, resulting in the reduction of the four-dimensional gauge group down to the commutant $\cH$ of $\cG \subset E_8$.
To preserve supersymmetry, the gauge field should satisfy the Hermitian Yang-Mills equations
\beq \label{HYM}
F_{\alpha \beta}= F_{\bar \alpha \bar \beta} = 0 \; ; ~g^{\alpha \bar \beta}F_{\alpha \bar{\beta}}  = 0 \ ,  
\eeq
where $F$ is the associated field strength. 
Although these equations can not be solved analytically, the Donaldson-Uhlenbeck-Yau theorem \cite{Donaldson:1985zz, UY} states that on a holomorphic (poly-)stable bundle, there exists a unique connection that solves Eq.~\eqref{HYM}.

So each of the heterotic vacua comes in two pieces: a Calabi-Yau three-fold $X$ and a holomorphic stable vector bundle $V$ thereon.  
Studying the detailed geometry, however, is not an easy task. To begin with, we do not even know the Ricci--flat metrics on Calabi-Yau three-folds. 
Fortunately, as will be seen shortly, it turns out that the topology of a vacuum already determines many interesting features of the four-dimensional effective theory. 

In order for the heterotic models to be ``Standard-like'', they must give rise to the correct gauge group, $SU(3)_C \times SU(2)_L \times U(1)_Y$, possibly with an extra $U(1)_{B-L}$ factor, as well as a correct spectrum for light particles coming in three generations. 
Firstly, the choices $\cG = SU(4)$ and $SU(5)$ for the structure group of $V$, reduce the $E_8$ to the four-dimensional gauge groups $\cH = SO(10)$ and $SU(5)$, respectively, which are desirable in the viewpoint of Grand Unification.\footnote{The choice $\cG=SU(3)$ also gives rise to $E_6$ GUT models. However, they have an inherent trouble in doublet-triplet splitting of Higgs multiplet (see, for a recent example, Ref.~\cite{Anderson:2009mh}) and hence, we will not address the models of this type here.}
The light particles then arise from the branching of the adjoint $\bf 248$ of $E_8$ into $\cG \times \cH$, and the spectrum is determined by various bundle-valued cohomologies on the Calabi-Yau three-fold \cite{GSW}, as summarised in Table~\ref{t:spectrum}. 
Of course, the gauge group $\cH$ should be further broken down to a Standard-like one, and discrete Wilson-lines are made use of, if there ever exists any, for this second breaking. 

\begin{table}[thb] 
{
\renewcommand{\arraystretch}{0.8}
{\begin{center}
\begin{tabular}{|c||c|c|c|}\hline 
\small{$\mathcal G$} &\small{$\mathcal H$} & \small{Branching of $\bf 248$ under $\mathcal G \times \mathcal H \subset E_8$} &  \small{Particle Spectrum} \\ \hline \hline

{}

\footnotesize{$SU(4)$} & \footnotesize{$SO(10)$} & \footnotesize{$({\bf 1}, {\bf 45}) \oplus ({\bf 4}, {\bf 16}) \oplus ({\bf \overline 4}, {\bf \overline {16}}) \oplus ({\bf 6}, {\bf 10}) \oplus ({\bf 15}, {\bf 1})$} & 
\footnotesize{
$
\ba{rcl}
n_{16}&=&h^{1}(X,V)\\
n_{\overline{16}}&=&h^{1}(X,V^\star)=h^2(V)\\
n_{10}&=&h^{1}(X,\wedge ^{2}V)\\
n_{1}&=&h^{1}(X,V\otimes V^\star)
\ea
$
}\\  \hline
\footnotesize{$SU(5)$} & \footnotesize{$SU(5)$} & \footnotesize{$({\bf 1}, {\bf 24}) \oplus ({\bf 5}, {\bf 10}) \oplus (\overline {\bf 5}, \overline{\bf 10}) \oplus ({\bf 10}, \overline{\bf 5}) \oplus (\overline{\bf 10}, {\bf 5}) \oplus ({\bf 24}, {\bf 1})$} &
\footnotesize{
$
\ba{rcl}
n_{10}&=&h^{1}(X,V)\\ 
n_{\overline{10}}&=&h^{1}(X,V^\star)=h^2(V)\\ 
n_{5}&=&h^{1}(X,\wedge^{2}V^\star)\\
n_{\overline{5}}&=&h^{1}(X,\wedge ^{2}V)\\
n_{1}&=&h^{1}(X,V\otimes V^\star)
\ea
$
}\\ \hline
\end{tabular}
\end{center}}
\vspace{-.15in}\hspace{.3in}\parbox{6.7in}
{\caption{\label{t:spectrum}\sf A vector bundle $V$ with structure group $\mathcal G=SU(4)$ or $SU(5)$, respectively, breaks the $E_8$ group of the heterotic string into the Grand Unifying group $\mathcal H=SO(10)$ or $SU(5)$. The four-dimensional low-energy representation contents of $\mathcal H$ arise from the branching of the $\bf 248$ adjoint of $E_8$ under $\mathcal G \times \mathcal H \subset E_8$. The particle spectrum is obtained as various bundle-valued cohomology groups. 
}}}
\vspace{-.25in}
\end{table}

In this paper, we shall make it clear how the construction of Standard-like heterotic vacua turns into the integer combinatorics for a discrete system.
Specifically, the Calabi-Yau three-folds will be {\it torically} constructed and described by the combinatorics of {\it reflexive} lattice polytopes\cite{Batyrev:1994hm}.\footnote{In algebraic geometry, Calabi-Yau three-folds are, in general, realised as complete intersections of hypersurfaces in toric varieties of dimension greater than or equal to four, but this paper will only be dealing with single--hypersurface cases.}  
Next, {\it monad} vector bundles~\cite{OSS} will be constructed thereon, equivalent of turning on internal gauge fluxes over the Calabi-Yau three-folds. 

The remainder of this paper is structured as follows.
In the ensuing two sections, we lay down the foundation by explaining the basic mathematical toolkit for describing $\cN=1$ heterotic vacua. 
Next, in section~\ref{sec4}, further constraints will be imposed on the internal geometry so that the resulting $\cN=1$ four-dimensional effective theory may mimic the Standard Model.
We shall conclude in section~\ref{sec5} with a summary and outlook. 

\section{Toric Construction of Calabi-Yau three-folds} \label{sec2}
Soon after the famous 7890 Calabi-Yau three-folds were realised as complete intersections of hypersurfaces in multi-projective spaces \cite{Candelas:1987du,Candelas:1987kf, Green:1987cr, He:1990pg,Gagnon:1994ek}, Kreuzer and Skarke have classified the Calabi-Yau three-folds that arise as codimension--one hypersurfaces in toric four-folds, comprising a much bigger dataset \cite{Kreuzer:1995cd, Kreuzer:2000xy, Kreuzer:2006ax}. 
This construction, first proposed by Batyrev \cite{Batyrev:1994hm}, involves an extensive usage of toric geometry.  
Here we do not intend by any means to give a pedagogical introduction to toric geometry.
The readers interested in the details of this subject are referred either to the maths texts \cite{fulton, oda, cox:review, textbook-toric} or to the excellent, introductory reviews for physicists \cite{Greene:1996cy, Bouchard:2007ik}. 

\subsection{Ambient Toric four-folds}
The Calabi-Yau three-folds $X$ are embedded in toric four-folds $\cA$ as hypersurfaces and therefore, we shall start with the description of these ambient toric varieties. 
A toric four-fold is described by the combinatorial data called a {\it fan} in $\IR^4$, which is a collection of convex {\it cones} in $\IR^4$ with their common apex at the origin $O=(0,0,0,0)$. 
For the sake of Calabi-Yau sub-varieties, however, every fan is not appropriate. 
We first define a certain class of convex {\it polytopes} in $\IR^4$, of which fans of a special kind are made. 

The polytopes considered here must contain the origin $O$ as the unique interior lattice point and all the vertices must lie in the lattice $\IZ^4 \subset \IR^4$.
Such polytopes are called {\it reflexive}. 
It can be shown that for a given reflexive polytope $\Delta$ in $\IR^4$, the {\it dual} polytope $\Delta^\circ$ defined by 
\beq
\Delta^\circ = \{ \bold{v} \in {\IR^4} ~|~ \left\langle \bold{m},\bold{v} \right\rangle \geq -1\;\;{\forall}\, \bold{m} \in \Delta \} \
\eeq
also has all its vertices on the lattice $\IZ^4$, like the original polytope $\Delta$ does.
To this dual polytope $\Delta^\circ$ we can associate a collection of the convex cones over all its faces, forming the fan for our toric four-fold $\cA$.  

Now, as for the construction of toric four-fold from a given fan in $\IR^4$, several equivalent methods are known. 
What best suits our purpose amongst them is Cox's homogenous-coordinate approach~\cite{Cox:1993fz}, where a complex homogeneous coordinate $x_{\rho}$ is associated to each one-dimensional cone $\rho$ in the fan. 
Thus, if the fan has $k$ edges, there are $k$ homogeneous coordinates $(x_1, \cdots, x_k)$ for $\mathbb{C}^k$.
The next task is to identify a certain measure zero subset $Z$ of $\mathbb{C}^k$ which should be removed. Let $S$ be a set of edges that do not span any cone in the fan and let $Z(S) \subset \IC^k$ be the linear subspace defined by setting $x_{\rho} = 0\;,\; \forall\, \rho \in S$. 
Now let $Z \subset \IC^k$ be the union of the subspaces $Z(S)$ for all such $S$. 
Then the toric four-fold is constructed as a quotient of $\IC^k - Z$ by the following $(\IC^*)^{k-4}$ action: 
\beq \label{homo}
(x_1, \cdots, x_k) \sim (\lambda_r^{\beta^r_{~1}} x_1, \cdots, \lambda_r^{\beta^r_{~k}} x_k) \ , ~\lambda_r \in \IC^* \text{~for~} r=1, \cdots, k-4 \ , 
\eeq
where the coefficients $\beta^r_{~\rho}$ are defined by the linear relations $\sum \limits_{\rho=1}^{k} \beta^r_{~\rho} \bold v_\rho =0$ amongst the edges. Hence, $\beta^r_{~\rho}$ form a $(k-4) \times k$ matrix which is often referred to as a {\it charge matrix} \cite{horietal}.
The identification rule in Eq.~\eqref{homo} can be schematically written as  \beq \label{quotient} \cA = (\IC^k-Z)/(\IC^*)^{k-4} \ . \eeq
Note that the construction of toric four-folds in Eq.~\eqref{quotient} naturally generalises that of projective space $\IP^4$, the simplest toric four-fold, in which case $Z = \{O\}$ and $k=5$, that is,
\beq \label{P4} \IP^4 = (\IC^5-\{O\})/\IC^* \ . \eeq

\subsection{Calabi-Yau three-folds} \label{subsec2.2}
A Calabi-Yau hypersurface $X$ to the toric four-fold $\cA$ is constructed in a straight-forward manner without requiring any further data:
as long as the polytope $\Delta$ is {reflexive} it also defines~$X \subset \cA$.
Note that in this case, $\Delta^\circ$ is also a reflexive polytope since $(\Delta^\circ)^\circ = \Delta$. 
To a reflexive polytope $\Delta$ in $\IR^4$, we can associate a family of Calabi-Yau three-folds $X$ defined as the vanishing loci of the polynomials of the form
\begin{equation}
\label{deq}
P_{\{C_{\bold m} \}}(x_1 , \cdots, x_k ) =\sum\limits_{\bold{m} \in \Delta} C_{\bold{m}} \prod\limits_{\rho =
1}^k x_\rho^{\langle \bold{m}, \bold{v_\rho} \rangle + 1} \ ,
\end{equation}
where $x_{\rho = 1, \ldots, k}$ are the $k$ homogeneous coordinates of $\cA$ associated to the lattice vertices $\bold v_{\rho=1, \dots, k}$ of $\Delta^\circ$, and $C_{\bold{m}}$ are numerical coefficients parameterising the complex structure of $X$.

Heterotic compactifications ask for compact Calabi-Yau three-folds that are smooth. 
However, a toric four-fold $\cA$ constructed by Eq.~\eqref{quotient} usually bears singularities and they in general descend to the hypersurfaces $X$ too. 
In order to make $X$ nonsingular, we partially de-singularise $\cA$ so that the hypersurfaces may avoid the singularities of the ambient space~\cite{Batyrev:1994hm}. 
This process corresponds to triangulating the (dual) polytope in a special way, and is called an {\it MPCP-triangulation}.\footnote{{MPCP} is a short for Maximal, Projective, Crepant and Partial. A triangulation is said to be {\it Maximal} if all lattice points of the polytope are involved, {\it Projective} if the K\"{a}hler cone of the resolved manifold has a nonempty interior, and {\it Crepant} if no points outside the polytope are taken. In practice, all possible MPCP-triangulations of a given reflexive polytope are searched by the computer package PALP~\cite{Kreuzer:2002uu}.} 

As for the statistics, a total of $473,800,776$ reflexive polytopes in $\IR^4$ have been classified~\cite{Kreuzer:1995cd, Kreuzer:2000xy, Kreuzer:2006ax}, each of which gives rise to a toric four-fold $\cA$ as well as a family of Calabi-Yau three-folds $X$. 
It turns out that only $124$ out of them describe smooth manifolds, for which no MPCP-triangulations are required. 

\section{Monad Construction of Vector Bundles} \label{sec3}
In the physics literature, especially in the context of heterotic string phenomenology, construction of vector bundles have been attempted in several ways. 
They include {\it spectral cover construction}~\cite{Friedman:1997ih, Donagi:1998xe,Donagi:1999gc, Diaconescu:1998kg, Andreas:1999ty, Curio:2004pf, Andreas:2003zb}, {\it bundle extension}~\cite{Braun:2005zv, Braun:2005ux, Braun:2005bw} and the mixture thereof~\cite{Donagi:2004ub}. 
In many of them, it was essential for the base three-folds to have a torus-fibration structure. 
On the other hand, {\it monad construction}~\cite{OSS} does not assume any extra structure and has proved particularly useful for algorithmically scanning a vast number of vector bundles~\cite{Anderson:2007nc, Anderson:2008uw, He:2009wi, toricpic, Anderson:2009mh}.

A {\it monad} vector bundle is essentially the quotient of two Whitney sums of line bundles. 
More precisely, a monad bundle $V$ over a Calabi-Yau three-fold $X$ is defined by the short exact sequence of the form:
\be\label{monad}
0 \longrightarrow V  \longrightarrow \bigoplus\limits_{i=1}^{r_b} \cO_X(\bold{b}_i)  \stackrel{f}{\longrightarrow} \bigoplus\limits_{j=1}^{r_c} \cO_X(\bold{c}_j)  \longrightarrow 0 \ , 
\ee
where $\bold b_i$ and $\bold c_j$ are integer vectors of length $h^{1,1}(X)$, representing the first Chern classes of the summand line bundles $\cO_X(\bold{b}_i)$ and $\cO_X(\bold{c}_j)$. 
The bundle $V$ is a holomorphic $U(n)$-bundle, where 
\beq \label{rank} n=r_b- r_c
\eeq
is the rank of $V$.

From Eq.~\eqref{monad}, one can readily read off the Chern class of $V$:
\bea
\label{c}
\nn c_1 (V) &=& \left( \sum_{i=1}^{r_b} b^r_i - \sum_{j=1}^{r_c} c^r_j \right) J_r \ , \\
c_2(V) &=& \frac{1}{2} d_{rst} \left(\sum_{j=1}^{r_c} c^s_j c^t_j - \sum_{i=1}^{r_b} b^s_i b^t_i \right) \nu^r \ , \\
\nn c_3(V) &=& \frac{1}{3} d_{rst} 
   \left(\sum_{i=1}^{r_b} b^r_i b^s_i b^t_i - \sum_{j=1}^{r_c} c^r_j
   c^s_j c^t_j \right) \ ,
\label{chernV}
\eea
where $J_r \in H^{1,1}(X, \IR)$ represent the harmonic $(1,1)$-forms $c_1(\cO_X(\bold e_r))$,  the $d_{rst}$ are the triple intersection numbers defined by
\beq \label{intersec-def}
d_{rst} = \int_X J_r \wedge J_s \wedge J_t \ , 
\eeq
and the $\nu^r$ are the 4-forms furnishing the dual basis to the K\"ahler generators $J_r$, subject to the duality relation
\beq \label{nubasis}
\int_{X} J_r \wedge \nu^s = \delta_r^s.
\eeq 
As can be seen from Eq.~\eqref{c}, the Chern class of $V$ only depends on the integer parameters $\bold b_i$ and $\bold c_j$, as well as the topology of the base manifold $X$.
Choosing an appropriate morphism $f$ in the defining sequence Eq.~\eqref{monad} corresponds to the tuning of more refined invariants of $V$. 

\section{Towards the Standard Model} \label{sec4}
Section~\ref{sec2} and section~\ref{sec3} have shown that the vacuum topology is essentially described by lattice vertices and integer parameters, both of which are discrete and combinatorial in nature.
One can therefore attempt to construct $\cN=1$ heterotic vacua in an algorithmic way. 
Torically constructed Calabi-Yau three-folds form a dataset of reflexive polytopes represented by the lattice vertices, and monad bundles are explored on each of the base manifolds by varying the integer parameters. 

\subsection{Phenemenological Constraints on the Vacua}

With the geometric constraints so far explained, one would only be able to guarantee the right number of supersymmetry at low energy, that is, $\cN=1$ at $D=4$. 
Since the goal of string phenomenology is to obtain (supersymmetric versions of) the Standard Model, more criteria should further be imposed on the $\cN=1$ vacua.
To make things clear, let us emphasize that in this paper the terminology ``Standard-like'' model will imply the following:
\bi
\item Gauge invariance under $SU(3)_C \times SU(2)_L\times U(1)_Y$, possibly with an extra $U(1)_{B-L}$ factor;
\item Three generations of quarks and leptons, and no exotics;
\item Cancellation of heterotic anomaly.
\ei
Here we translate the above three phenomenological constraints into the conditions on the vacuum topology. 

\subsubsection{Gauge Group}
As explained in the Introduction, the structure group $\cG$ of the visible sector bundle sits in $E_8$ and its commutant $\cH$ becomes the low-energy gauge group. 
In order to obtain $\cH= SO(10)$ and $SU(5)$, one must choose $\cG=SU(4)$ and $SU(5)$, respectively. 
In particular, the rank of the bundle should be either $4$ or $5$, and hence, by Eq.~\eqref{rank}, 
\beq r_b - r_c = 4  \text{~or~} 5 \ , \eeq
where $r_b$ and $r_c$ are the ranks of the two vector bundles in the defining sequence Eq.~\eqref{monad} of $V$. 
What is more, since the structure group should be ``special'' unitary, the first Chern class $c_1(V)$ of $V$ is to vanish. 
By Eq.~\eqref{chernV}, this corresponds to 
\beq \sum_{i=1}^{r_b} \bold b_i = \sum_{j=1}^{r_c} \bold c_j \ ,  \eeq
where $\bold b_i = (b_i^1, \dots, b_i^{h^{1,1}})$ and $\bold c_j = (c_j^1, \dots, c_j^{h^{1,1}})$ are the $h^{1,1}$-tuples of integers labelling the summand line bundles, and hence, parameterising the monad $V$.

We still have to break the GUT group further down to a Standard-like one, and this second breaking will require $\pi_1(X)$-Wilson-lines. 
However, given the observation that most of the torically constructed Calabi-Yau three-folds have a trivial first fundamental group~\cite{Batyrev:2005jc}, they must be quotiented out by freely-acting discrete symmetries so that we may turn on appropriate Wilson-lines.
Therefore, we will eventually have to look for a discrete symmetry group $G$ that acts freely on $X$, and make a quotient space $\hat X = X/G$, which will then have a non-trivial first fundamental group $\pi_1(\hat X) \simeq G$.

\subsubsection{Cancellation of Heterotic Anomaly}
Heterotic models need to satisfy a well-known anomaly condition. 
So far we have only mentioned one holomorphic vector budle $V$ for the visible sector, but the theory has another bundle $\tilde V$ for the hidden sector. 
Heterotic vacua can also have five-branes whose strong-coupling origin is M5-branes.
In order to keep the four-dimensional Lorentz symmetry, their world-volumes must stretch along the external Minkowski $M_4$.
The remaining two dimensions should then wrap holomorphic two-cycles in $X$ for unbroken supersymmetry.
Thus, the homology classes associated with these two-cycles must be {\it effective} and hence, belong to the Mori cone in $H_2(X, \IZ)$. 
In other words, the corresponding four-fourms must belong to the corresponding cone in $H^4(X, \IZ)$. 

In this most general set-up, heterotic anomaly cancellation imposes a topological constraint relating the Calabi-Yau three-fold, the two vector bundles and the five-brane classes. 
When $c_1(TX) = c_1(V) = c_1(\tilde V) = 0$, the anomaly condition can be expressed, at the level of cohomology, as
\beq \label{c2}
c_2(TX) - c_2(V) - c_2(\tilde V) = W \ , 
\eeq
where $W= \sum\limits_i W_i$ is the sum of the five-brane classes. 
Note that $W$ itself should also belong to the Mori cone of $X$ as all the summands $W_i$ do. 
In our discussion, however, without mentioning the second bundle $\tilde V$ we presume a trivial bundle for the hidden sector.
Thus, the anomaly constraint in Eq.~\eqref{c2} says that $c_2(TX) - c_2(V)$ is effective.

\subsubsection{Particle Spectra}
Table~\ref{t:spectrum} shows how the low-energy particle spectra are determined from various bundle-valued cohomology groups. 
Assuming that $V$ is a stable bundle\footnote{Testing the bundle stability is indeed one of the crucial steps for our model construction. However, it is not at all an easy task to check if a given bundle is stable. So our strategy is first to make use of some consequences of stability and then to check the validity at the very end of the story. In this article, focusing on the discrete combinatorics for the vacua, we will not say more about the issue of stability.} implies that $h^0(X,V) =0 = h^3(X,V)$, and hence, to obtain three net-generations of quarks and leptons, we must have
\beq\label{index-thm}
-\frac{1}{2} \int_X c_3(V) = h^1(X,V)  - h^2(X,V) =3 |G| \ , 
\eeq
where the Aiyah-Singer index theorem~\cite{GnH} has been applied to the differential operator $\slashed{\partial}_V$ on $X$, and $|G|$ is the order of the discrete symmetry group $G$, with which we will have to quotient the ``upstairs'' three-folds $X$.

\subsection{Discrete System for Standard-like Vacua}
In this sub-section, we briefly summarise the model-building requirements that have so far been discussed:
\bi
\item {\bf Calabi-Yau three-fold}

A reflexive polytope $\Delta \subset \IR^4$ describes a Calabi-Yau three-fold $X$. 
In the computer package PALP~\cite{Kreuzer:2002uu}, by inserting the list of lattice vertices of $\Delta$, or equivalently, the corresponding ``weight system''~\cite{Kreuzer:2000xy}, one can obtain all the topological invariants of $X$ relevant to the heterotic compactification.
\item {\bf Monad vector bundle}

The $r_b+r_c$ integer vectors $\bold b_i$ and $\bold c_j$ of length $h^{1,1}(X)$, each labelling a line bundle summand, parameterise our monad bundle $V$. 

\item {\bf Standard-like constraints}

The internal backgrounds are also constrained by the Standard-like phenomenology.  
It turns out that given a Calabi-Yau three-fold $X$, that is, for a fixed topology of $X$, the monad parameters $\bold b_i$ and $\bold c_j$ must obey the algebraic equations of degree 0, 1, 2 and 3 shown in Table~\ref{t:BasicConstraints}. 
\begin{table}[t!b!h!]
{\renewcommand{\arraystretch}{1.8}
{\begin{center} \scriptsize
~~~~~~~~\begin{tabular}{|c||c|c|} \hline 
\text{Physics Origin} & \text{~~~~~Background Geometry~~~~~} & \text{Algebraic constraints on} $\bold b_i$ \text{and} $\bold c_j$ \\[4pt] \hline \hline 
\text{~~Gauge Group~~} &  
$
\ba{rcl}
&\text{a.}& {\cal G} = SU(n) ,~\text{for~}n=4,5 \\ 
&\text{b.}& X \text{ has a discrete free action } G 
\ea 
$ &
$
\ba{rcl}
&\text{a1.}& r_b - r_c = 4 \text{~or~} 5 \\ 
&\text{a2.}& \sum\limits_{i=1}^{r_b} b_i^r = \sum\limits_{j=1}^{r_c} c_j^r\ ,  ~\forall r 
\ea 
$
\\[17pt] 
\hline
{Anomaly} & $c_2(TX)-c_2(V)$ \text{is effective} &
$
c_2(TX)_r - \frac{1}{2}d_{rst} \left(\sum\limits_{j=1}^{r_c} c^s_j c^t_j - \sum\limits_{i=1}^{r_b} b^s_i b^t_i \right)
>0 \ , \forall r$
\\[6pt] 
\hline
\text{Particle Spectra} &
$
\ba{rcl}
\frac{1}{2} \int_X c_3(V) = - 3 |G| 
\ea
$ &
$
d_{rst} 
   \left(\sum\limits_{i=1}^{r_b} b^r_i b^s_i b^t_i - \sum\limits_{j=1}^{r_c} c^r_j
   c^s_j c^t_j \right) = -18 |G|
$\\[6pt] 
\hline 
\end{tabular}
\end{center}}
\hspace{.85in}\parbox{6in}{\caption{\label{t:BasicConstraints} \sf The list of Standard-like constraints on the $\cN=1$ heterotic vacua, each described by a Calabi-Yau three-folds $X$ and a $\cG$-bundle $V$ thereon; the integers $b_i^r$ and $c_j^r$ parameterise the bundle $V$ as in Eq.~\eqref{monad}, where $i=1, \dots, r_b$, $j=1, \dots, r_c$ and $r=1, \dots, h^{1,1}(X)$. The second column states the constraints on the background geometry and the third column expresses the corresponding algebraic equations that the parameters $b_i^r$ and $c_j^r$ must obey, where $d_{rst} = \int_X J_r \wedge J_s \wedge J_t$ are the intersection numbers for a given basis $\{J_r\}_{1 \leq r \leq h^{1,1}(X)}$ for $H^{1,1}(X,\IR)$.}}
}
\end{table}
\ei

Note that the integer combinatorics under the algebraic constraints of small degrees has formed a simple discrete system for Standard-like heterotic vacua. 
However, this system of vacua is far from being finite yet. 
To begin with, there are no upper bounds on $r_b$ and $r_c$ that count the number of monad parameters.
Before initiating an exploration of the landscape, one first needs to add more constraints to make the system finite, and those extra constraints had better be related to preferred phenomenology. 
Now, if the line bundle summands in Eq.~\eqref{monad} are ample, or equivalently, if all the monad parameters $b_i^r$ and $c_j^r$ are positive,\footnote{To be precise, Kodaira's vanishing assumes that the vectors $\bold b_i$ and $\bold c_j$ lie in the K\"{a}hler cone of $X$. In case the K\"{a}hler cone does not coincide with the positive region, one may redefine the standard basis vectors of $H^{1,1}(X, \IR)$ to be the K\"{a}hler cone generators. For this to work, however, the cone generators should form a linearly independent basis and hence, we implicitly restrict ourselves to the Calabi-Yau three-folds of this type.} then by Kodaira's vanishing theorem~\cite{GnH, HA}, the cohomology group $H^2(X, V)$ vanishes and hence, the low-energy effective theory acquires no anti-generations.
Of course this is a phenomenologically preferred feature, although not necessary.
We call a monad {\it positive} if it is only parameterised by positive integers, and {\it semi-positive} if all its parameters are either positive or zero.
Secondly, one can also constrain the relative size of these monad parameters so that  each entry of the vector $\bold b_i - \bold c_j$ may be non-negative, for all $i$ and $j$. 
In this case, we will call the monad {\it generic} since the monad map $f$ in Eq.~\eqref{monad}, thought of as an $r_c \times  r_b$ matrix of polynomials, may generically have all the entries nonzero. 

\subsection{Exploring a Region of the Landscape}
As for the first step, one can think of exploring generic, positive monads over a ``small'' class of Calabi-Yau three-folds.
As was mentioned in sub-section~\ref{subsec2.2}, the total dataset of torically constructed Calabi-Yau three-folds are way too large to grasp altogether. 
Therefore, at the initial stage, those in ``smooth'' ambient spaces have first been considred amongst the total of 500 million~\cite{He:2009wi}. 
It turns out that over these $O(100)$ manifolds, the generic, positive monads are finite in number under the constraints in Table~\ref{t:BasicConstraints}, and the Standard-like vacua have indeed been classified, resulting in 61 candidate models. 

Based on this experience, one can become more ambitious and extend the vacuum search, both bundle-wise and Calabi-Yau-wise. 
Firstly, with the positivity condition a bit relaxed, the generic,  semi-positive monads have been explored over the same $O(100)$ Calabi-Yau three-folds~\cite{He:2009wi}.
The Standard-like vacua with the monads of this type turn out to form an infinite class and hence, they have been explored under an artificial upper bound on the monad parameters, resulting in 85 models.
Secondly, the programme has also been extended to include singular ambient manifolds with small $h^{1,1}$~\cite{toricpic}.
A total of $O(300)$ torically constructed Calabi-Yau three-folds have the Hodge number $h^{1,1} \leq 3$, and the generic, positive monads have been classified thereon, giving rise to new candidate models.

\section{Summary and Outlook} \label{sec5}
In this paper, we have discussed a systematic approach towards Standard-like heterotic vacua.
The proposed algorithms have indeed been implemented in a computer package~\cite{CICYpackage}. 
Simplicity of the integer combinatorics for the $\cN=1$ heterotic vacua was the essential feature that made this approach a tractable programme.
It was motivated by the general observation that any carefully-chosen single model is likely to fail the detailed structure of the Standard Model. 
Thus, the spirit of the programme is to obtain a large number of Standard-like models, on which further constraints should be imposed later on to refine the set of candidates, eventually reaching the ``genuin'' Standard Model(s).

The combinatorics of toric geometry has been invaluable for constructing toric ambient four-folds, to which Calabi-Yau three-folds have been embedded as hypersurfaces, and for computing their topological invariants relevant to the four-dimensional
phenomenology. 
Smooth ambient four-folds have been considered as a starter~\cite{He:2009wi}, and general ambient four-folds have also been dealt with~\cite{toricpic} by partially resolving the singularities, if they bear any, so that the smoothness of the hypersurface Calabi-Yau
three-folds are guaranteed.
In both cases, the generic, positive monads (and some semi-positive ones, too, in the former case) have been probed under the Standard-like criteria.
We have thus obtained a set of canidate models, that are anomaly-free and that have a chance to produce three generations of quarks and leptons without any anti-generations. 

To guarantee the three-generation property of these candidates, further study of discrete symmetries of the manifolds are essential.
Braun has recently classified the free group actions on complete intersection Calabi-Yau three-folds in multi-projective spaces~\cite{Braun:2010hr} and his algorithm can in principle be generalised to toric cases. 
The line-bundle cohomologies on the torically constructed Calabi-Yau three-folds are also an essential part of the model building.
The starting point would be to work out the cohomologies on the ambient toric varieties, which have already been investigated in the mathematics and physics literatures~\cite{textbook-toric, Rahn:2010fm, Blumenhagen:2010pv, Blumenhagen:2010ed}.
Practical conversion of this information into the line-bundle cohomologies on the hypersurfaces is a rewarding work along the line of monad bundles and heterotic strings.
As for the completion of the detailed particle spectra, the cohomologies of the monads in different representations are also to be revealed.

\section*{Acknowledgement}
The author is grateful to Yang-Hui~He and Andre~Lukas for collaborations, and to Lara~Anderson and James~Gray for discussion, on the projects which this paper is based upon. 
He would especially like to thank Maximilian~Kreuzer, who passed away in the middle of collaborations on part of the work reviewed here, for invaluable correspondence and advice.




\begin{thebibliography}{99}

\bibitem{Candelas:1985en}
P.~Candelas, G.~T.~Horowitz, A.~Strominger and E.~Witten,
``Vacuum Configurations for Superstrings,''
Nucl.\ Phys.\  B {\bf 258} (1985) 46.

\bibitem{GSW}
M.~Green, J.~Schwarz and E.~Witten, 
\emph{Superstring Theory, vol.2}, 
Cambridge Monographs on Mathematical Physics, Cambridge University Press, Cambridge, UK, 1987. 

\bibitem{Witten:1996mz}
E.~Witten,
``Strong Coupling Expansion of Calabi-Yau Compactification,''
Nucl.\ Phys.\  B {\bf 471} (1996) 135
[arXiv:hep-th/9602070].

\bibitem{Lukas:1997fg}
A.~Lukas, B.~A.~Ovrut and D.~Waldram,
``On the Four-Dimensional Effective Action of Strongly Coupled Heterotic   String Theory,''
Nucl.\ Phys.\  B {\bf 532} (1998) 43
[arXiv:hep-th/9710208].

\bibitem{Lukas:1998yy}
A.~Lukas, B.~A.~Ovrut, K.~S.~Stelle and D.~Waldram,
``The Universe as a Domain Wall,''
Phys.\ Rev.\  D {\bf 59} (1999) 086001
[arXiv:hep-th/9803235].

\bibitem{Lukas:1998hk}
A.~Lukas, B.~A.~Ovrut and D.~Waldram,
``Non-Standard Embedding and Five-Branes in Heterotic M-Theory,''
Phys.\ Rev.\  D {\bf 59} (1999) 106005
[arXiv:hep-th/9808101].

\bibitem{Lukas:1998tt}
A.~Lukas, B.~A.~Ovrut, K.~S.~Stelle and D.~Waldram,
``Heterotic M-Theory in Five Dimensions,''
Nucl.\ Phys.\  B {\bf 552} (1999) 246
[arXiv:hep-th/9806051].

\bibitem{Donaldson:1985zz}
S.~K.~Donaldson,
``Anti Self-Dual Yang-Mills Connections over Complex Algebraic Surfaces and  Stable Vector Bundles,''
Proc.\ Lond.\ Math.\ Soc.\  {\bf 50} (1985) 1.

\bibitem{UY}
K.~Uhlenbeck, S.~T.~Yau, 
``On the existence of hermitian-yang-mills connections in stable vector bundles,''
Comm. Pure Appl. Math. {\bf 39} (1986) S257.

\bibitem{Batyrev:1994hm}
V.~V.~Batyrev,
``Dual Polyhedra and Mirror Symmetry for Calabi-Yau Hypersurfaces in Toric   Varieties,''
J.\ Alg.\ Geom.\  {\bf 3} (1994) 493.

\bibitem{OSS}
C.~Okonek, M.~Schneider, H.~Spindler,
\emph{Vector Bundles on Complex Projective Spaces},
Birkh\"{a}user, 1980.

\bibitem{Anderson:2007nc}
L.~B.~Anderson, Y.~H.~He and A.~Lukas,
``Heterotic Compactification, an Algorithmic Approach,''
JHEP {\bf 0707} (2007) 049
[arXiv:hep-th/0702210].

\bibitem{Anderson:2008uw}
L.~B.~Anderson, Y.~H.~He and A.~Lukas,
``Monad Bundles in Heterotic String Compactifications,''
JHEP {\bf 0807} (2008) 104
[arXiv:0805.2875 [hep-th]].

\bibitem{Anderson:2009mh}
L.~B.~Anderson, J.~Gray, Y.~H.~He and A.~Lukas,
``Exploring Positive Monad Bundles and a New Heterotic Standard Model,''
JHEP {\bf 1002} (2010) 054
[arXiv:0911.1569 [hep-th]].

\bibitem{He:2009wi}
Y.~H.~He, S.~J.~Lee and A.~Lukas,
``Heterotic Models from Vector Bundles on Toric Calabi-Yau Manifolds,''
JHEP {\bf 1005} (2010) 071
[arXiv:0911.0865 [hep-th]].

\bibitem{toricpic}
Y.~H.~He, M.~Kreuzer, S.~J.~Lee, A.~Lukas,
In preparation.

\bibitem{CICYpackage}
A.~Lukas, L.~Anderson, J.~Gray, Y.~H.~He, S.~J.~Lee,
\emph{CICY package},
based on methods described in Refs.~\cite{Anderson:2009mh, Anderson:2007nc, Anderson:2008uw, He:2009wi}.

\bibitem{Candelas:1987kf}
  P.~Candelas, A.~M.~Dale, C.~A.~Lutken and R.~Schimmrigk,
  ``Complete Intersection Calabi-Yau Manifolds,''
  Nucl.\ Phys.\  B {\bf 298}, 493 (1988).

\bibitem{Candelas:1987du}
  P.~Candelas, C.~A.~Lutken and R.~Schimmrigk,
``Complete Intersection Calabi-Yau Manifolds 2. Tree Generation Manifolds,''
  Nucl.\ Phys.\  B {\bf 306}, 113 (1988).

\bibitem{Green:1987cr}
  P.~S.~Green, T.~Hubsch and C.~A.~Lutken,
 ``All Hodge Numbers Of All Complete Intersection Calabi-Yau Manifolds,''
  Class.\ Quant.\ Grav.\  {\bf 6}, 105 (1989).

\bibitem{He:1990pg}
  A.~m.~He and P.~Candelas,
``On the Number of Complete Intersection Calabi-Yau Manifolds,''
  Commun.\ Math.\ Phys.\  {\bf 135}, 193 (1990).

\bibitem{Gagnon:1994ek}
  M.~Gagnon and Q.~Ho-Kim,
``An Exhaustive list of complete intersection Calabi-Yau manifolds,''
  Mod.\ Phys.\ Lett.\  A {\bf 9}, 2235 (1994).

\bibitem{Kreuzer:1995cd}
M.~Kreuzer and H.~Skarke,
``On the Classification of Reflexive Polyhedra,''
Commun.\ Math.\ Phys.\  {\bf 185} (1997) 495
[arXiv:hep-th/9512204].

\bibitem{Kreuzer:2000xy}
M.~Kreuzer and H.~Skarke,
``Complete Classification of Reflexive Polyhedra in Four Dimensions,''
Adv.\ Theor.\ Math.\ Phys.\  {\bf 4} (2002) 1209
[arXiv:hep-th/0002240].

\bibitem{Kreuzer:2006ax}
M.~Kreuzer,
``Toric Geometry and Calabi-Yau Compactifications,''
Ukr.\ J.\ Phys.\  {\bf 55} (2010) 613
[arXiv:hep-th/0612307].

\bibitem{fulton}
W.~Fulton,
\emph{Introduction to Toric Varieties},
Princeton University Press, 1993.

\bibitem{oda}
T.~Oda,
\emph{Convex Bodies and Algebraic Geometry},
Springer-Verlag, 1988.

\bibitem{cox:review}
D.~Cox,
``Recent Developments in Toric Geometry'',
[arXiv:alg-geom/9606016v1].

\bibitem{textbook-toric}
D.~Cox, J.~Little and H.~Schenck,
\emph{Toric Varieties},
AMS, to be published in 2011. 

\bibitem{Greene:1996cy}
B.~R.~Greene,
``String Theory on Calabi-Yau Manifolds,''
arXiv:hep-th/9702155.

\bibitem{Bouchard:2007ik}
V.~Bouchard,
``Lectures on Complex Geometry, Calabi-Yau Manifolds and Toric Geometry,''
arXiv:hep-th/0702063.

\bibitem{Cox:1993fz}
D.~Cox,
``The Homogeneous Coordinate Ring of a Toric Variety, Revised Version,''
arXiv:alg-geom/9210008.

\bibitem{horietal}
K.~Hori~et~al., 
\emph{Mirror Symmetry}, 
American Mathematical Society, 2003.


\bibitem{Kreuzer:2002uu}
M.~Kreuzer and H.~Skarke,
``Palp: a Package for Analyzing Lattice Polytopes with Applications to Toric Geometry,''
Comput.\ Phys.\ Commun.\ {\bf 157} (2004) 87
[arXiv:math/0204356].

\bibitem{Friedman:1997ih}
  R.~Friedman, J.~W.~Morgan, E.~Witten,
 ``Vector bundles over elliptic fibrations,''
  [alg-geom/9709029].
  
\bibitem{Donagi:1998xe}
  R.~Donagi, A.~Lukas, B.~A.~Ovrut, D.~Waldram,
  ``Nonperturbative vacua and particle physics in M theory,''
  JHEP {\bf 9905}, 018 (1999).
  [hep-th/9811168].

\bibitem{Donagi:1999gc}
  R.~Donagi, A.~Lukas, B.~A.~Ovrut, D.~Waldram,
  ``Holomorphic vector bundles and nonperturbative vacua in M theory,''
  JHEP {\bf 9906}, 034 (1999).
  [hep-th/9901009].
  
\bibitem{Diaconescu:1998kg}
  D.~-E.~Diaconescu, G.~Ionesei,
  ``Spectral covers, charged matter and bundle cohomology,''
  JHEP {\bf 9812}, 001 (1998).
  [hep-th/9811129].
  
\bibitem{Andreas:1999ty}
  B.~Andreas, G.~Curio, A.~Klemm,
  ``Towards the Standard Model spectrum from elliptic Calabi-Yau,''
  Int.\ J.\ Mod.\ Phys.\  {\bf A19}, 1987 (2004).
  [hep-th/9903052].
  
\bibitem{Andreas:2003zb}
  B.~Andreas, D.~Hernandez-Ruiperez,
  ``Comments on N = 1 heterotic string vacua,''
  Adv.\ Theor.\ Math.\ Phys.\  {\bf 7}, 751-786 (2004).
  [hep-th/0305123].
  
\bibitem{Curio:2004pf}
  G.~Curio,
  ``Standard model bundles of the heterotic string,''
  Int.\ J.\ Mod.\ Phys.\  {\bf A21}, 1261-1282 (2006).
  [hep-th/0412182].
  
\bibitem{Donagi:2004ub}
  R.~Donagi, Y.~-H.~He, B.~A.~Ovrut, R.~Reinbacher,
  ``The Spectra of heterotic standard model vacua,''
  JHEP {\bf 0506}, 070 (2005).
  [hep-th/0411156].
  
\bibitem{Braun:2005ux}
  V.~Braun, Y.~-H.~He, B.~A.~Ovrut, T.~Pantev,
  ``A Heterotic standard model,''
  Phys.\ Lett.\  {\bf B618}, 252-258 (2005).
  [hep-th/0501070].
  
\bibitem{Braun:2005bw}
  V.~Braun, Y.~-H.~He, B.~A.~Ovrut, T.~Pantev,
  ``A Standard model from the E(8) x E(8) heterotic superstring,''
  JHEP {\bf 0506}, 039 (2005).
  [hep-th/0502155].
  
\bibitem{Braun:2005zv}
  V.~Braun, Y.~-H.~He, B.~A.~Ovrut, T.~Pantev,
  ``Vector bundle extensions, sheaf cohomology, and the heterotic standard model,''
  Adv.\ Theor.\ Math.\ Phys.\  {\bf 10}, 4 (2006).
  [hep-th/0505041].
  
\bibitem{Kachru:1995em}
  S.~Kachru,
  ``Some three generation (0,2) Calabi-Yau models,''
  Phys.\ Lett.\  {\bf B349}, 76-82 (1995).
  [hep-th/9501131].
  
\bibitem{Anderson:2007nc}
  L.~B.~Anderson, Y.~-H.~He, A.~Lukas,
  ``Heterotic Compactification, An Algorithmic Approach,''
  JHEP {\bf 0707}, 049 (2007).
  [hep-th/0702210 [HEP-TH]].
  
\bibitem{Anderson:2008uw}
  L.~B.~Anderson, Y.~-H.~He, A.~Lukas,
  ``Monad Bundles in Heterotic String Compactifications,''
  JHEP {\bf 0807}, 104 (2008).
  [arXiv:0805.2875 [hep-th]].
  
\bibitem{Anderson:2009mh}
  L.~B.~Anderson, J.~Gray, Y.~-H.~He, A.~Lukas,
  ``Exploring Positive Monad Bundles And A New Heterotic Standard Model,''
  JHEP {\bf 1002}, 054 (2010).
  [arXiv:0911.1569 [hep-th]].
 
  
\bibitem{Batyrev:2005jc}
V.~Batyrev and M.~Kreuzer,
``Integral Cohomology and Mirror Symmetry for Calabi-Yau 3-Folds,''
arXiv:math/0505432.

\bibitem{GnH}
P.~Griffiths and J.~Harris, 
{\em Principles of algebraic geometry,}
Wiley, 1978.

\bibitem{HA}
R.~Hartshorne, 
{\em Algebraic Geometry,}
Springer, GTM52, Springer-Verlag, 1977.

\bibitem{Braun:2010hr}
V.~Braun,
``Discrete Wilson Lines in F-Theory,''
arXiv:1010.2520 [hep-th].

\bibitem{Blumenhagen:2010pv}
R.~Blumenhagen, B.~Jurke, T.~Rahn and H.~Roschy,
``Cohomology of Line Bundles: a Computational Algorithm,''
J.\ Math.\ Phys.\ {\bf 51} (2010) 103525
[arXiv:1003.5217 [hep-th]].

\bibitem{Rahn:2010fm}
T.~Rahn and H.~Roschy,
``Cohomology of Line Bundles: Proof of the Algorithm,''
J.\ Math.\ Phys.\ {\bf 51} (2010) 103520
[arXiv:1006.2392 [hep-th]].

\bibitem{Blumenhagen:2010ed}
R.~Blumenhagen, B.~Jurke, T.~Rahn and H.~Roschy,
``Cohomology of Line Bundles: Applications,''
arXiv:1010.3717 [hep-th].


\end{thebibliography}
\end{document}